\UseRawInputEncoding
\documentclass[aps,preprint,llncs]{revtex4}
\usepackage[utf8]{inputenc}
\usepackage[english]{babel}
\usepackage[T1]{fontenc}
\usepackage{amsmath}
\usepackage{amsfonts}
\usepackage{amssymb}
\usepackage{xcolor}
\usepackage[colorlinks=true,linkcolor=blue,filecolor=blue, urlcolor=blue,citecolor=blue]{hyperref}
\usepackage{graphicx}
\usepackage{float}
\usepackage{color}
\usepackage{soul}
\usepackage{siunitx}
\usepackage{physics}
\usepackage{lineno}
\usepackage{subfiles}

\newpage

\begin{document}

\title{ 
Incommensurate Moiré Stacking and Landau Quantization Without External Magnetic Field in Turbostratic Graphene}
\author{Mona Garg$^{1}$}
\email{mona@iisermohali.ac.in}
\author{Ankit Kumar$^2$}
\author{Deepti Rana$^1$}
\author{Anmol Arya$^1$}
\author{Aswini R$^1$}
\author{Umesh Waghmare$^2$}
\author{G. U. Kulkarni$^2$}
\author{Goutam Sheet$^1$}

\email{goutam@iisermohali.ac.in}
\affiliation{$^1$Indian Institute of Science Education and Research (IISER) Mohali, Sector 81, S. A. S. Nagar, Manauli, PO 140306, India}
\affiliation{$^2$Jawaharlal Nehru Centre for Advanced Scientific Research (JNCASR), Jakkur, Bengaluru, Karnataka 560064, India}

\begin{abstract}

Turbostratic multilayer graphene, composed of randomly twisted and stacked graphene sheets, offers a naturally disordered yet tunable platform for exploring moiré physics beyond tedious artificial stacking. Using scanning tunneling microscopy/spectroscopy (STM/STS) and Raman analysis, we uncover a wide distribution of twist angles and stacking configurations spontaneously formed across large-area turbostratic films. In several regions, we identify overlapping incommensurate moiré patterns consistent with locally chiral trilayer stacking. We observe van Hove singularities and reconstructed Dirac-like spectra whose angle dependence supports strong interlayer electronic coherence. In the highly strained trilayered regions, we observe peaks in the local density-of-states with characteristic scaling of the quantized Landau levels strikingly even in the absence of a magnetic field. They arise from the strain-induced pseudo-magnetic fields ($\sim$ 26 T), making turbostratic graphene a single natural platform to explore the physics of moiré structures as well as of the pseudo-electromagnetic fields.

\end{abstract}


 \maketitle


Moiré superlattices in two-dimensional (2D) van der Waals materials have emerged as a fertile platform for exploring strongly correlated quantum phenomena, including unconventional superconductivity, correlated insulating states, and the quantum anomalous Hall effect\cite{luican_single-layer_2011, ohta_evidence_2012, dean_hofstadters_2013, stauber_chiral_2018, cao_correlated_2018, cao_unconventional_2018, chen_evidence_2019, sharpe_emergent_2019, chen_tunable_2020, serlin_intrinsic_2020, park_tunable_2021,hao_electric_2021,kim_evidence_2022,zhang_promotion_2022, mukherjee_superconducting_2025}. These phases often arise in twisted bilayer and trilayer graphene systems, where precise control of interlayer twist angles enables fine-tuning of the electronic band structure\cite{bistritzer_moire_2011, phong_band_2021}. However, the experimental realization of precisely twist-angle-controlled systems remains technically demanding, particularly when a wide range of angles or large-area samples are desired. Turbostratic graphene is a naturally occurring form of multilayer graphene characterized by rotational misalignment between adjacent layers and it provides a promising alternative\cite{mogera_highly_2015}. Unlike artificially engineered twisted bilayers obtained through exfoliation-based techniques, turbostratic graphene intrinsically hosts a broad distribution of twist angles and stacking configurations within a single continuous film\cite{lenski_raman_2011,kim_raman_2012,garlow_large-area_2016, sun_hetero-site_2021,kokmat_growth_2023}. This offers a platform for the spontaneous realization of diverse moiré periodicities and emergent electronic structures without the need for tedious and expensive fabrication.

While twisted bilayer graphene has been extensively explored over the past decade, recent research has increasingly turned toward multilayer graphene systems, particularly twisted trilayer graphene (TTG), which is formed by stacking three graphene layers with specific interlayer twist angles. Such a system is characterized by two independent twist angles, $\theta_{12}$ and $\theta_{23}$, corresponding to the relative rotations between layers 1–2 and 2–3, respectively. In the special case where $\theta_{12} = -\theta_{23}$, the structure is mirror-symmetric and exhibits a single moiré periodicity due to commensurate interference. However, in the more general and less explored asymmetric configuration where $\theta_{12} \neq -\theta_{23}$, the system gives rise to two distinct moiré patterns, originating from independent interference between adjacent layer pairs \cite{mora_flatbands_2019, zhu_twisted_2020, zhang_correlated_2021, nakatsuji_multiscale_2023, yang_multi-moire_2024}. These moiré lattices are typically incommensurate, resulting in a quasi-crystalline stacking order and rich electronic structure. Recent experiments have demonstrated that such asymmetric trilayer configurations can also host flat bands and a superconducting phase under specific conditions \cite{uri_superconductivity_2023, hao_robust_2024}, further highlighting their physical significance. In this work, using low-temperature scanning tunneling microscopy and spectroscopy (STM/STS), we demonstrate that turbostratic graphene spontaneously forms regions with two misaligned moiré periodicities, providing direct evidence of chiral incommensurate trilayer stacking. Spectroscopic measurements in these regions reveal prominent features in the local density of states (LDOS), including pseudo–Landau level quantization, arising from strain-induced gauge fields. These features signal coherent electronic coupling across layers and confirm the emergence of nontrivial moiré stacking even in the absence of deterministic twist angle control. 
Our findings demonstrate that turbostratic graphene, typically considered electronically decoupled, hosts spatially resolved, coherent low-energy electronic states shaped by interlayer twist alignments, incommensurability and strain fields. These results establish turbostratic multilayers as a naturally disordered, yet theoretically rich, setting for exploring Moiré-driven emergent physics.

The turbostratic graphene films are grown by a Joule heating technique, which spontaneously forms wrinkles in the film. The high quality of the turbostratic films is evident from the detailed characterization of these films using Raman spectroscopy, electron diffraction, and electrical transport studies reported elsewhere\cite{mogera_intrinsic_2017, mogera_twisted_2018, gupta_twist-dependent_2020}. The details of experimental methods are provided in the supporting information (SI) file.
We emphasize that large-area regions with well-defined twist angles can, in principle, be identified, extracted, and transferred onto desired substrates for potential device applications\cite{mogera_intrinsic_2017, mogera_twisted_2018, gupta_twist-dependent_2020,gupta_ultrafast_2022, gupta_highly_2025}. 

\begin{figure}[h!]
    \includegraphics[width = 0.95\linewidth]{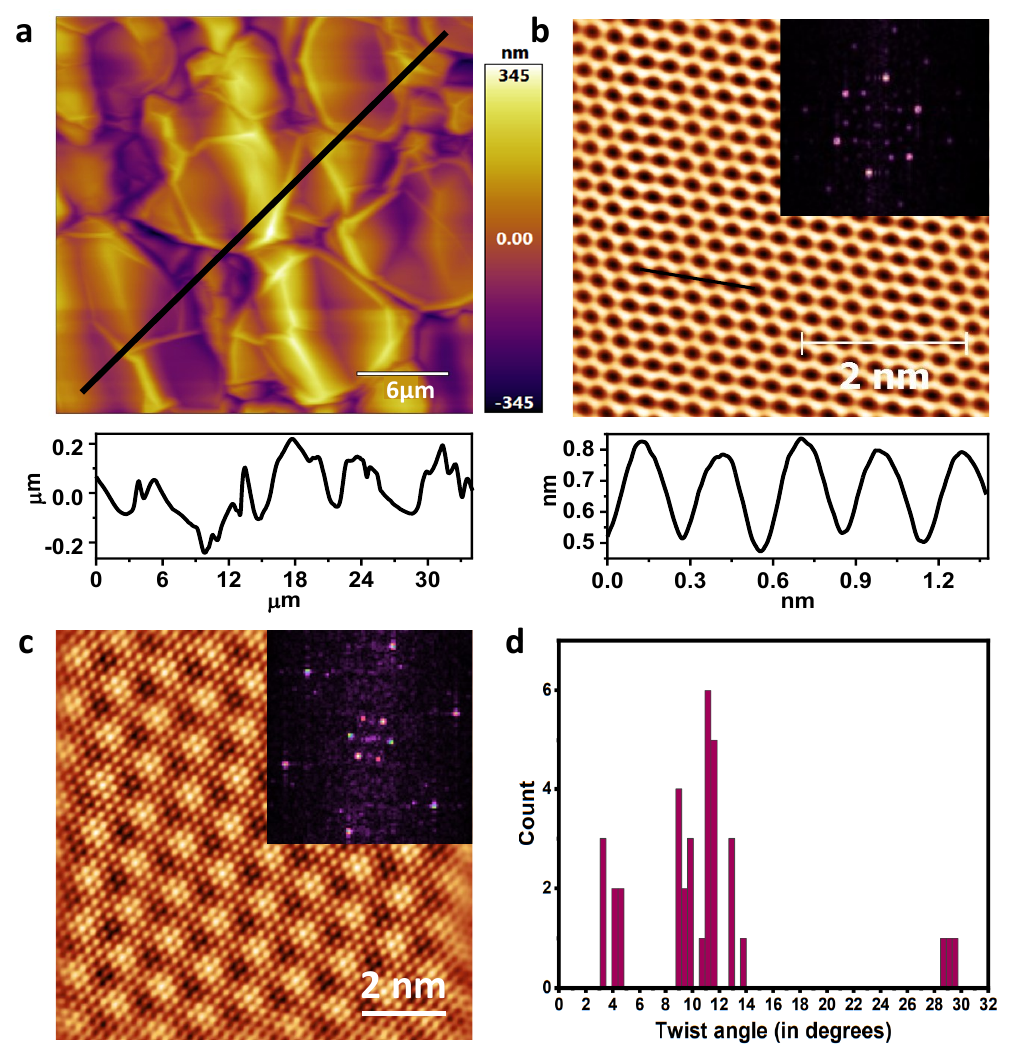}
    \caption {(a) AFM topograph showing wrinkles on the turbostratic graphene (above) and height profile across the marked region (below). (b) The atomically resolved image was taken on a 5 $\times$ 5 nm$^2$ area showing the graphene lattice (above) and height profile across the marked region (below). (c) STM topograph taken over an area of 10.7 $\times$ 10.7 nm$^2$ of turbostratic graphene showing moiré pattern with twist angle $\approx$ 12$^\circ$. (d) Distribution of different twist angles (in degrees) as obtained from the analysis of the moiré patterns. The $insets$ in (b) and (d) show the Fast Fourier Transform of the corresponding topographic images.}
\end{figure}

For characterization of the surface of the turbostratic graphene films, we first performed atomic force microscopy experiments at room temperature. Figure 1(a) presents an atomic force microscopy (AFM) topography that illustrates the presence of wrinkles in the film. By examining the line profile, as shown in the lower panel of Figure 1(a), it is evident that the wrinkles exhibit greater height compared to the flat regions of the film. In earlier reports, the wrinkled regions were found to be more conductive than the flat regions\cite{moun_enhanced_2021}. To probe the atomic scale microscopic structure of such wrinkled regions, we performed scanning tunneling microscopy and spectroscopy (STM/S) at 77 K. A small area STM topograph of 5$\times$ 5 nm$^2$ (Figure 1 (b)) in a selected region shows the presence of a lattice with a lattice constant matching that of single layer graphene (lower panel of Figure 1 (b)), indicating that these regions have top graphene layer significantly decoupled from the underlying stacking layers with high turbostraticity and that effectively behaves like suspended monolayer graphene\cite{mogera_highly_2015}. We found that there are different regions with varying degrees of turbostraticity and stacking patterns over the film surface, as shown in a representative moiré lattice image in Figure (1c). 

We note that the spatial extent of the ``coupled” regions with well-resolved moiré patterns was, on average, of the order $\sim$1 $\mu$m. This is confirmed by Raman mapping analysis as well as STM imaging over a large area. In the SI file, in Figures S1-S4, we have shown uniformity of the moiré patterns over large areas in different regions of the turbostratic film. The Raman mapping analysis is summarized in Figures S5 and S6 in SI file, which supports the large area uniformity of the moiré pattern in the turbostratic film. A large number of STM topographs were recorded in different regions and corresponding twist angles were calculated. The statistics of the twist angles obtained, as shown in Figure 1(d), indicate a wide range of twist angles present in the turbostratic graphene films. The distribution also reveals that the major fraction of the turbostratic film hosts a twist angle of around 12$^o$. The distribution of the twist angle discussed above is also supported by our Raman spectroscopy analysis conducted on various regions of the films, as presented in the SI file Figure S7.

\begin{figure}[h!]
    \begin{center}
    \includegraphics[width = 1\linewidth]{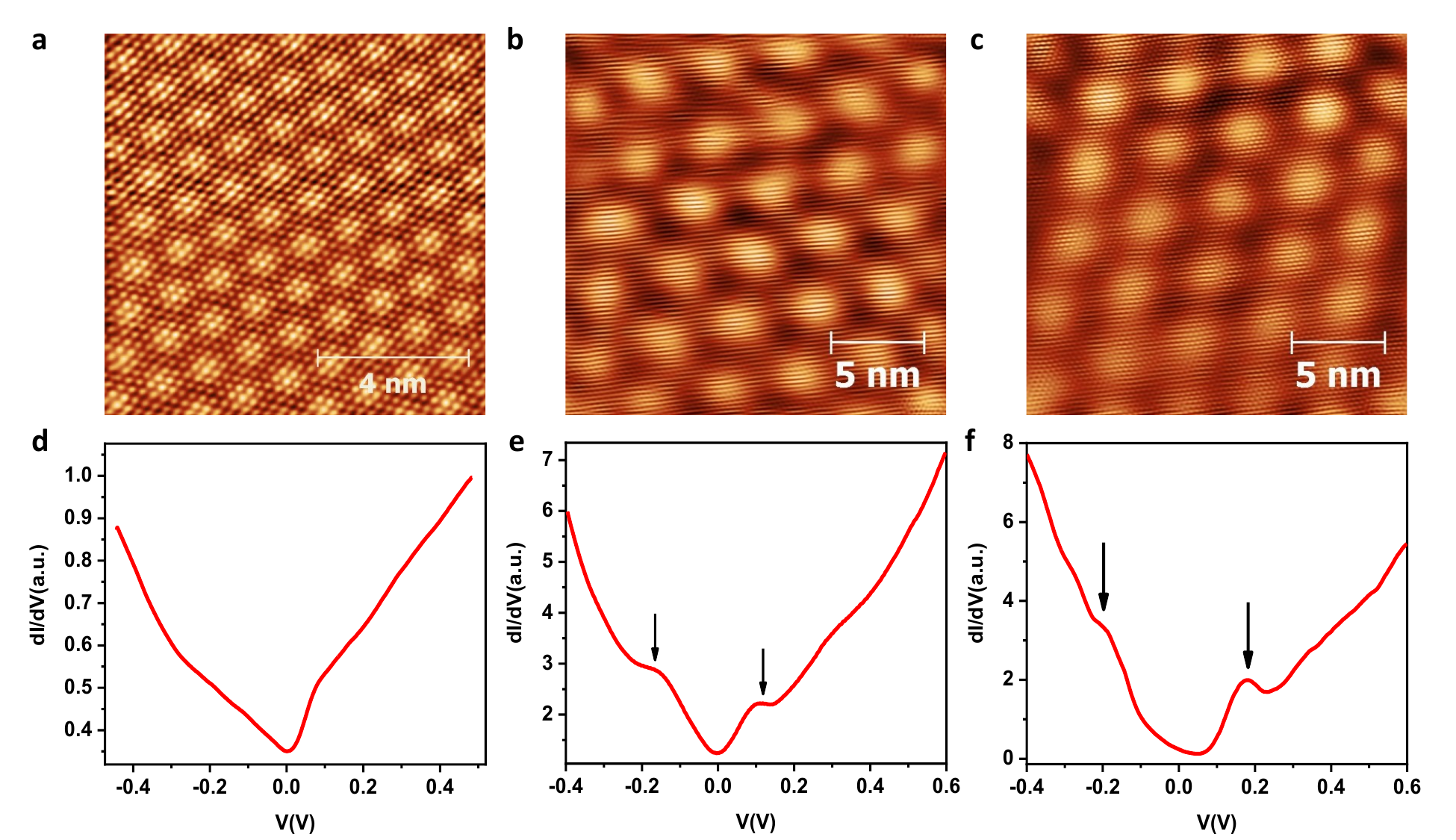}
    \end{center}
    \caption {(a) STM topograph taken over an area of 10 $\times$ 10 nm$^2$ of turbostratic graphene showing moiré pattern along with atomic resolution due to rotational twist among stacked layers. (b, c) Scan over 20 $\times$ 20 nm$^2$ area in different regions showing moiré pattern with different twist angles. (d-f) STS was performed at 77 K in different regions as shown in (a-c), respectively. Each $dI/dV$ plot is an average of 4 spectra. The black arrows in (e, f) show the position of vHSs.}
\end{figure}

To explore the local electronic states among different twisted-layer configurations, we performed scanning tunneling spectroscopy (STS) at various points on the surface of the turbostratic graphene films. Figure 2(a) depicts a 10$\times$ 10 nm$^2$ area showing a moiré pattern with an inter-layer twist angle of $\sim$ 14.7$^o$. The spectroscopic measurements in the same region display a Dirac cone-like linear energy dependence of the local density of states (Figure 2(d)). The existence of such a Dirac cone feature is known to arise for twisted graphene with large twist angles (>5$^o$). In such cases, the Dirac cones corresponding to each layer are shifted in the momentum space with mutual overlapping, leading to a reconstructed band structure which retains the linear dispersion for certain large twist angles\cite{latil_massless_2007, hass_why_2008, lopesdossantos_graphene_2007, luican_single-layer_2011, shallcross_electronic_2010}. For smaller twist angles, saddle point features appear in the band structure due to finite interlayer coupling. They appear as van Hove singularities (vHS)\cite{li_observation_2010, yan_angle-dependent_2012, brihuega_unraveling_2012, yan_angle-dependent_2014} in local density of states (LDOS) spectra. In Figure 2(b,c), we show the moiré pattern of two different regions with twist angles of 3.4$^{\circ}$ and 3.5$^{\circ}$, respectively. For such low twist angles, our STS spectra indeed revealed the vHS (Figure 2(e,f)). In these cases, we also found the evidence of angle-dependent variation of the energy gap $\sim$ 0.28 eV and $\sim$ 0.37 eV as shown in Figure 2(e,f), respectively.

\begin{figure}[h!]
    \begin{center}
    \includegraphics[width = 1\linewidth]{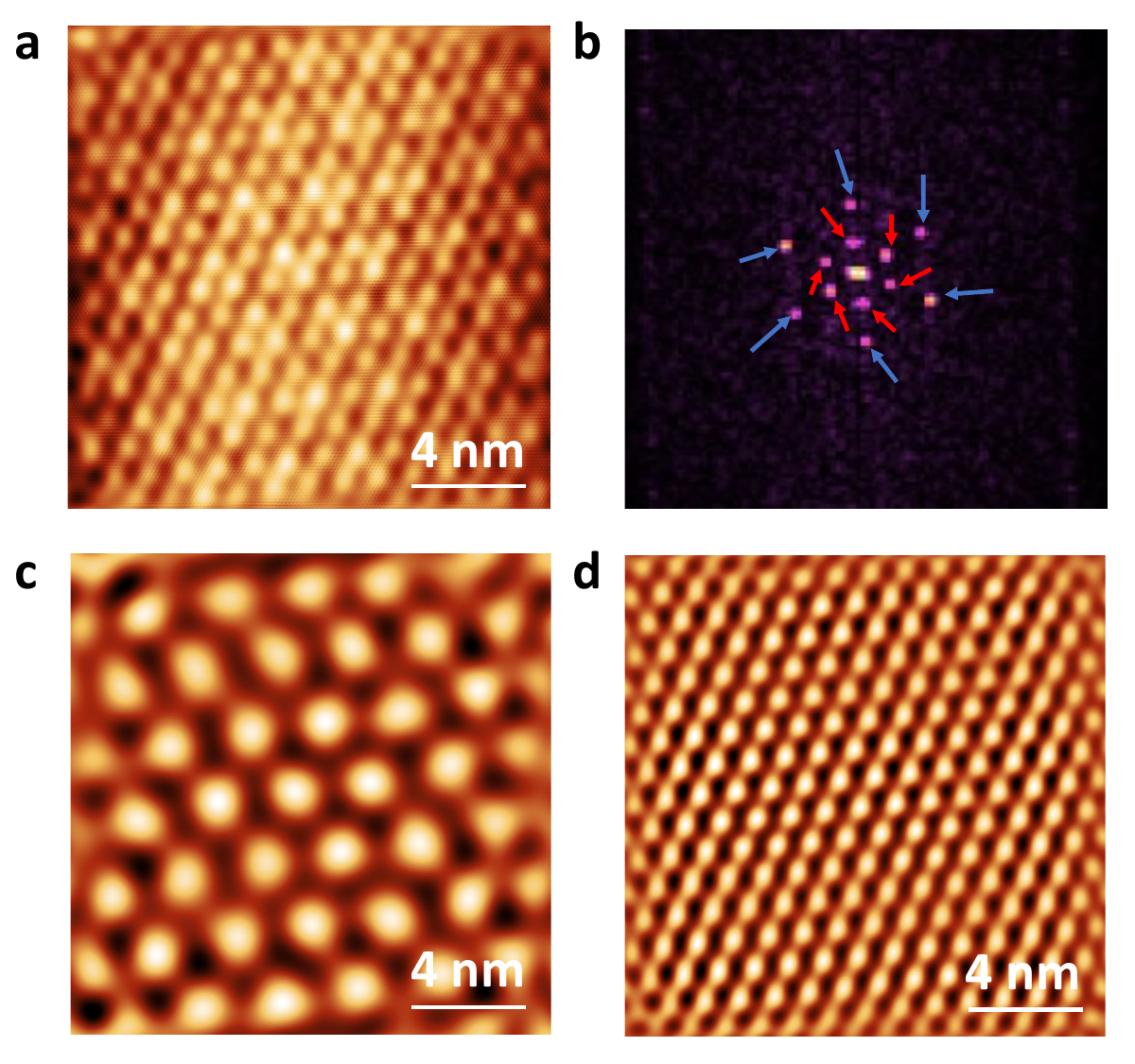}
    \end{center}
    \caption {(a) An STM topograph of 19 $\times$ 19 nm$^2$ area. (b) Fast Fourier Transform of the topographic image shown in (a). The red and blue arrows show the points corresponding to two different moiré periodicities. (c,d) The moiré image corresponding to the respective red and blue arrows in (b).}
\end{figure}

The atomic resolution image of another selected region over 19$\times$ 19 nm$^2$ area is shown in Figure 3(a). The image shows a graphene lattice as well as a complex pattern of bright spots (Moiré) with a larger lattice constant. To understand the origin of such a complex moiré pattern, we analyzed the image digitally. First, we performed a Fast Fourier Transform(FFT) of the topographic image, shown in Figure 3(b) and then sequentially masked different regions (indicated by arrows) of the FFT image to uncover the contribution of structures with different periodicity. Figure 3(c) displays the inverse FFT of the inner regions marked by red arrows, while Figure 3(d) shows the inverse FFT corresponding to the outer regions marked by blue arrows. These figures illustrate two types of moiré patterns characterized by different lattice constants. The twist angle between consecutive layers can be calculated using the relation $D_{ij}=\frac{a}{2sin(\frac{\theta_{ij}}{2})}$; where $D_{ij}$ is the observed moiré period between layers $i$ and $j$, $\theta_{ij}$ is the corresponding twist angle and $a$ is the monolayer graphene lattice constant \cite{hermann_periodic_2012}. Using this scheme, twist angles of around $\theta_{12} = 4.46^{\circ}$ and $\theta_{23} = 10^{\circ}$ were obtained for Figures 2(d) and 2(e), respectively. The observation of two distinct moiré patterns suggests two distinctly twisted interfaces within a local trilayer stacking. As it will be presented later, the scaling behaviour of a strain-induced pseudomagnetic field in the system supports the presence of trilayer stacking. 

We emphasize on the fact that the two periodicities coexist in real space rather than arising from separate domains or higher-order harmonics. Two moiré patterns with periods $D_{12}$ and $D_{23}$ are commensurate only if the ratio $\frac{D_{12}}{D_{23}} = \frac{p}{q}$, where $p$ and $q$ are small integers such that their corresponding reciprocal vectors satisfy the linear relation $\sum_{i=1}^{3} m_i \, \mathbf{g}^{(12)}_i + \sum_{i=1}^{3} n_i \mathbf{g}^{(23)}_i = 0$, where $\mathbf{g}^{(12)}_i$ and $\mathbf{g}^{(23)}_i$ are the $i$-th reciprocal lattice vectors of the moiré patterns generated by the (1,2) and (2,3) layer interfaces, respectively, and $m_i, n_i$ are integers. A nontrivial integer solution corresponds to a common reciprocal lattice vector, indicating commensurability; the absence of such a solution implies incommensurability. In our case, $D_{12} = 3.16$ nm and $D_{23} = 1.41$ nm. These give $\frac{D_{12}}{D_{23}} \approx 2.241$, which is not a rational number (not a ratio of two small integers) within experimental uncertainty, so no common superlattice exists and the stacking is incommensurate. This coexistence of two independent moiré modulations within the same region is therefore consistent with effective locally chiral, incommensurate trilayer stacking.

\begin{figure}[h!]
    \begin{center}
    \includegraphics[width = 1\linewidth]{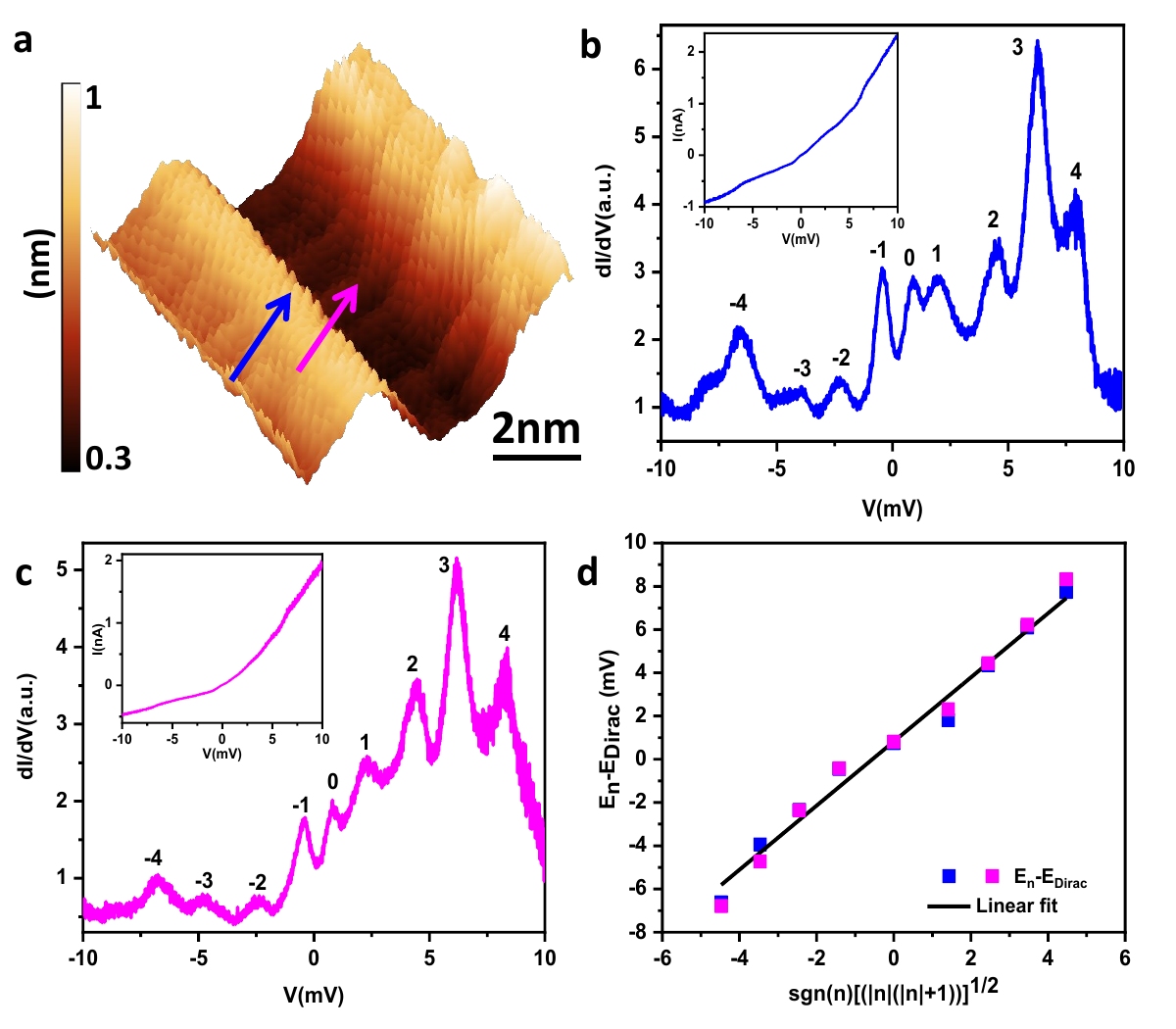}
    \end{center}
    \caption {(a) 10 $\times$ 10 nm$^2$ topography scan showing wrinkled region. The arrows show the tip position where scanning tunneling spectroscopy (STS) was done. (b, c) A series of peaks can be seen in the LDOS at a lower energy scale for STS performed at 350 mK. The inset shows the corresponding $I$-$V$ characteristics. (d) Scaling of E$_n$-E$_{Dirac}$ with $sgn(n)\sqrt{(|n|(|n|+1))}$.}
\end{figure}

Since the turbostratic graphene films also host a distribution of strain on the surface, we performed experiments down to 400 mK to investigate the effect of the local strain field on the electronic properties. For such experiments, we first obtained an atomically resolved topographic image of a highly strained area hosting a wrinkle (Figure 4(a)). The region over the 10 $\times$10 nm$^2$ area shows a prominent wrinkle with height $h$ $\sim$ 0.2 nm and width $w$ $\sim$ 3.5 nm (estimated from the FWHM of the line profile across the wrinkle). The underlying graphene lattice is visible, but is highly distorted and therefore the stacking order could not be resolved. The STM topograph recorded at a distance of $\sim$ 150 nm from the strained region (topograph in Figure 4 (a)) reveals the stacking of at least three layers of graphene with incommensurate stacking supported by an irrational $D_{12}/D_{23}$ ratio (Figure S8 in the SI file). Then, we performed local spectroscopy at various points in the same region. In Figures 4(b) and 4(c), we present two representative $dI/dV$ vs $V$ spectra corresponding to the two points indicated by the blue and magenta arrows in Figure 4(a). As seen in Figure 4(b,c), a series of peaks were observed in the LDOS plot as a function of energy. The corresponding $I$-$V$ characteristics are shown in the respective $inset$. The separation between consecutive peaks varies from $\sim$ 1.4 meV to $\sim$ 1.8 meV. These resemble the pseudo-Landau levels (PLL) induced by the coupling of the strain field with the relativistic charge carriers in graphene\cite{mao_evidence_2020, lee_quantum_2011}. Starting from the charge neutrality point, we labeled the PLL peaks in terms of $n$ where $n$ is an integer. We then investigated the scaling of the energy ($E_n$) of the $n$th peak with index $n$. As shown in Figure 4(d), we found that $E_n$ - $E_{Dirac}$ vary linearly with $sgn(n)\sqrt{(|n|(|n|+1))}$. Such a scaling behavior is known to exist for trilayer graphene \cite{yin_landau_2015}, and therefore is relevant in this case. This supports the idea that electronic wavefunctions are coherent across layers. Following the scheme presented elsewhere \cite{yin_landau_2015, levy_strain-induced_2010}, the expected strain-induced flux per wrinkle in distorted graphene is $\Phi$ = ($\beta$$h^2$/$l$$a$) $\Phi_0$, where $h$ and $l$ is the height and width of the wrinkle respectively, $a$ is the C-C bond length (0.142 nm), $\Phi_0$ is the quantum of flux and $\beta$ relates the change in the hopping amplitude between nearest neighbor carbon atoms to bond length. It has a typical magnitude of 2 $\leq$ $\beta$ $\leq$ 3 for graphene. This gives an estimate of the pseudomagnetic field $\sim$ 26 Tesla. This corresponds to magnetic length, $l_B$ $\sim$ 25 nm/$\sqrt{B(Tesla)}$ $\sim$ 4.8 nm. This is consistent with the length scale of the wrinkled region. The presence of high curvature of the wrinkles and the consequent large local strain gives rise to the high local pseudomagnetic field, making it possible to observe PLLs. This is consistent with the typical magnitude of PLL seen in certain 2D systems with linear bands\cite{kamboj_generation_2019}.

Finally, we also performed a temperature dependence study of the conductance spectra, displaying PLL in another region, as shown in Figure S9 of the SI file. The LDOS spectra at 400 mK clearly show a Dirac cone-like sharp dip around zero and Landau-level-like oscillations at higher bias (Figure S9(b)). With slowly rising temperatures upto 1.85 K, we found that while the Dirac cone feature remains locked at a given energy, everywhere else the oscillations gradually smear out (Figure S9(d)). This behavior over a rather narrow temperature range (between 400 mK and 1.85 K) indicates a low Dingle temperature, perhaps due to disorder in the turbostratic graphene films.


In summary, we have combined atomic-resolution STM imaging and tunneling spectroscopy to uncover the coexistence of two distinct moiré superlattices in multiple regions of a large area turbostratic trilayer graphene, arising from independent twist angles at the (1,2) and (2,3) interfaces. Fourier analysis reveals two independent sets of reciprocal lattice vectors whose real-space periods have an irrational ratio within experimental uncertainty, providing direct structural evidence of incommensurate stacking. Filtered inverse FFT reconstructions confirm that both periodicities coexist within the same spatial region, ruling out alternative explanations such as separate domains or higher-order harmonics.
In highly strained regions, we observe pronounced peaks in the local density of states, whose energies follow the characteristic scaling of quantized Landau levels for massive Dirac fermions in trilayer Bernal graphene. These findings present a natural platform  that enables exploration of reconstructed Dirac band physics, emergent gauge fields, and potentially correlated electronic states, all without artificial stacking or external magnetic fields.




We thank Prof. Tobias Stauber, ICMD-CSIC Madrid, for fruitful discussions. M.G. thanks the Council of Scientific and Industrial Research (CSIR), Government of India, for financial support through a research fellowship (Award No. 09/947(0227)/2019-EMR-I). D.R. acknowledges the Department of Science and Technology, Government of India for the INSPIRE Fellowship. G.S. acknowledges financial assistance from the Science and Engineering Research Board (SERB), Govt. of India (grant number: CRG/2021/006395).

The data that support the findings of this study are available within the article and the supporting information file.

The authors declare no competing financial interests.
\bibliography{main}

\pagebreak

\title{Supporting Information for \\
\large Incommensurate Moiré Stacking and Landau Quantization Without External Magnetic Field in Turbostratic Graphene}

\author{Mona Garg$^{1}$}
\email{mona@iisermohali.ac.in}
\author{Ankit Kumar$^2$, Deepti Rana$^{1}$, Anmol Arya$^1$, Aswini R$^1$, Umesh Waghmare$^2$, G. U. Kulkarni$^2$} 
\author{Goutam Sheet$^1$}
\email{goutam@iisermohali.ac.in}

\affiliation{$^1$Indian Institute of Science Education and Research (IISER) Mohali, Sector 81, S. A. S. Nagar, Manauli, PO 140306, India}

\affiliation{$^2$Jawaharlal Nehru Centre for Advanced Scientific Research (JNCASR), Jakkur, Bengaluru, Karnataka 560064, India}

\maketitle

\section*{Materials and methods}
To grow a turbostratic graphene film on Nickel (Ni), a
A current-carrying polycrystalline nickel foil was connected to electrodes and a solution of naphthalene in chloroform was applied to the Nickel foil by dropping it onto the surface. The nickel foil was then subjected to Joule heating, reaching a high temperature of approximately 800$^\circ$C, using a direct current (DC) source with a current density of 120 $A/mm^{2}$. This process lasted for 15 minutes under a pressure of 4 mTorr. After heating, the nickel foil was rapidly cooled. The graphene film thus formed shows a wrinkled structure which is believed to develop during the cooling phase after the high-temperature synthesis process, primarily due to the differential contraction of the Ni substrate and graphene.


Atomic force microscopy imaging was performed at ambient (MFP-3D) of Asylum Research system), in non-contact mode using a highly doped Si-cantilever. The resonance frequency of the AFM probe was around 75 kHz.

STM/S measurements were performed using an ultra-low temperature and ultra-high vacuum UNISOKU USM 1300 equipped with RHK R9 controller. Before STM measurements, the sample of size  4 mm $×\times$ 2 mm was mounted on an Arsenic-doped Silicon chip using silver epoxy. The sample was heated $in$-$situ$ in the UHV preparation chamber by flowing current through the Si-chip to remove any oxide layer/surface contamination and then transferred to the scanning stage. Similarly, the electrochemically etched Tungsten tip was cleaned inside UHV by heating treatment. STM/S measurements were performed at 77 K and 400 mK as mentioned in the main text. All the topography imaging was done in constant current mode. Gwyddion software was used for performing the Fast Fourier transform.

The Raman spectroscopy measurements were performed using an XploRA Plus (HORIBA) Raman microscope. A 532 nm laser served as the excitation source. Spectral data were acquired using a 100$\times$ objective with a numerical aperture of 0.8. Raman mapping analysis was performed using LabSpec 6 software.

\section*{Uniform Moiré lattice over a large regions}

\begin{figure}[h!]
    
    \includegraphics[width = 1\linewidth]{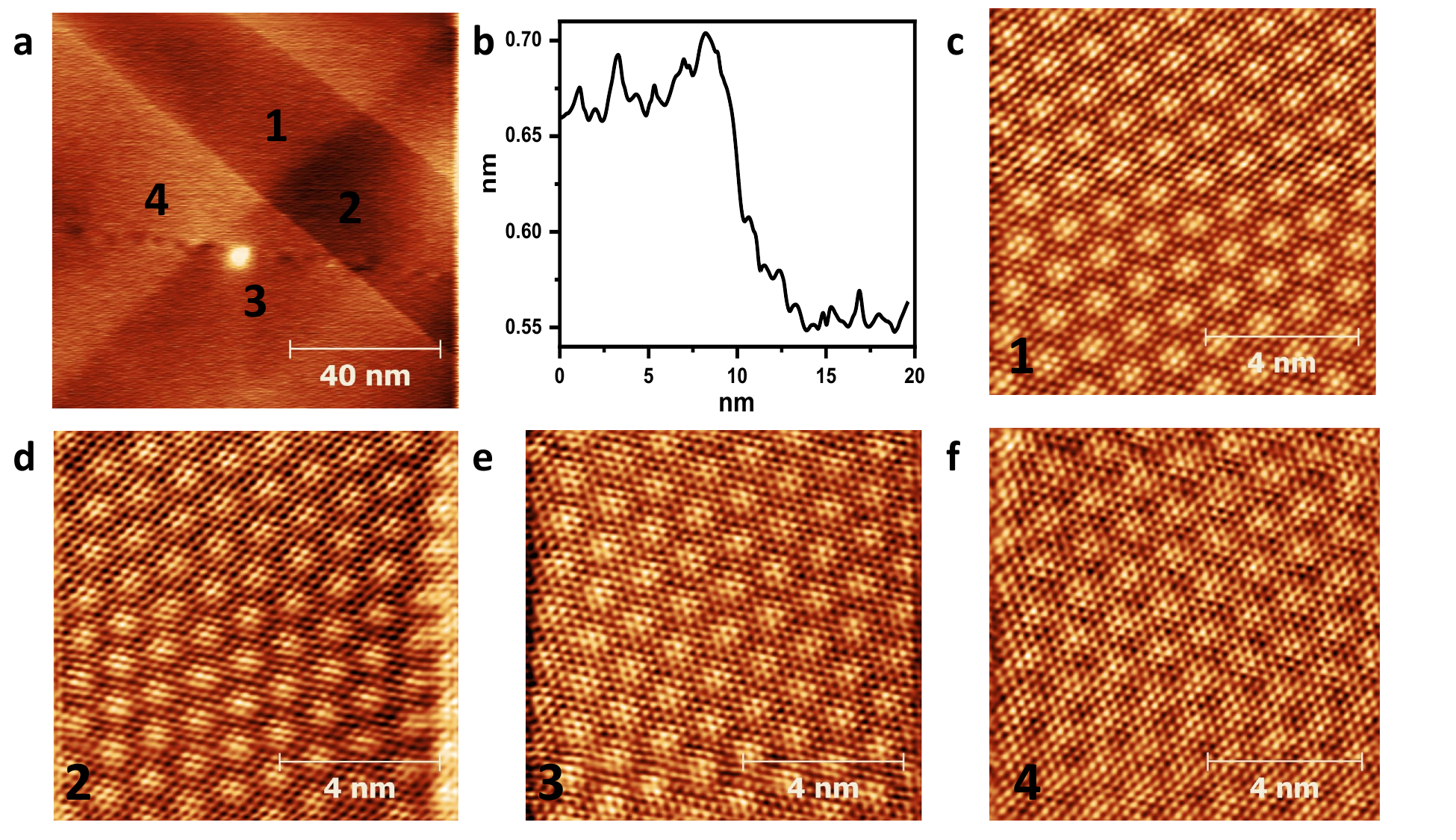}
    
    \caption {(a)Large area topograph of size 107 $\times$ 107 nm$^2$ showing wrinkles and a bright defect (sample bias = 400 mV, tunneling current = 450 pA). (b) Height profile across the marked region in (a). (c-f) Atomic resolution images showing the moiré pattern in different regions of (a).}
\end{figure}

In Figure S1(a), we show an STM image over a large area of 107 $\times$ 107 nm$^2$ recorded at 77 K. This area captures two wrinkles almost normal to each other. The wrinkled regions span over a lateral length scale of $\sim$ 30 nm. Atomic resolution imaging reveals a distinct Moiré pattern. A line profile over the wrinkles shows a variation of $\sim$ 0.1 nm across the wrinkle (Figure S1(b)). In this region, the area on top of the wrinkles (Figure S1(c,d)) and the area around them (Figure S1(e,f)) display the same moiré pattern, indicating that the wrinkled as well as the flat regions have identical inter-layer stacking and twist angles. 

The maximum scan area that can be scanned without coarse movement in our STM is approximately 1 $\mu$m. The following STM topograph in Figures S2-S4 were analyzed in different regions to confirm the uniformity of the twist angles. The coordinates of the acquired topograph (in nm) from the center of the scan window of 1 $\mu$m $\times$ 1 $\mu$m are noted and provided for each topograph.
\begin{figure}[h!]
    
    \includegraphics[width= 0.9\linewidth]{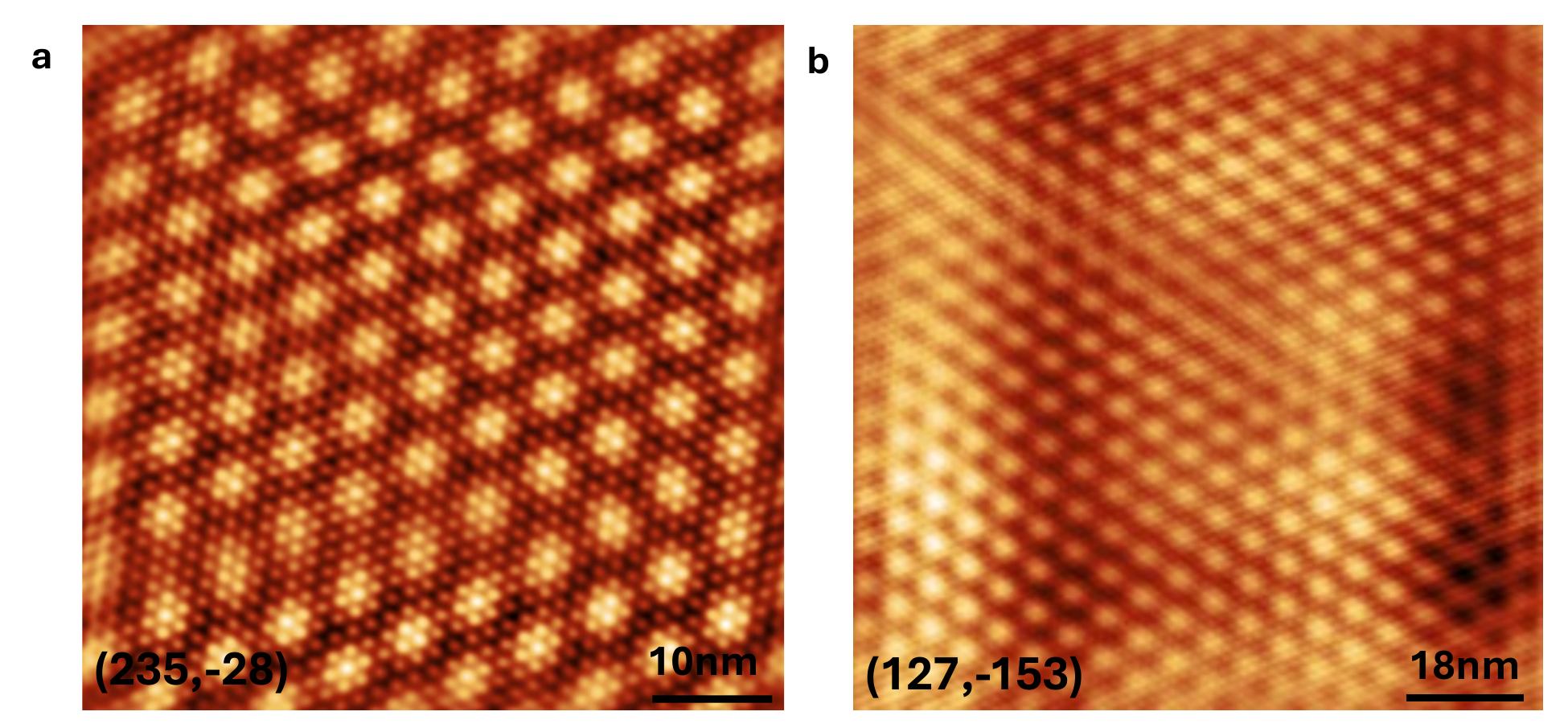}
    
    \caption {(a) An STM topograph of a 53.8 $\times$ 53.8 nm$^2$ area taken at specified coordinates, showing two different moiré periodicities corresponding to twist angles of approximately 11.2$^\circ$ and 3.1$^\circ$. (b) Another STM topograph of a 90 $\times$ 90 nm$^2$ area, approximately 100 nm away from (a), also showing the same moiré twist angles.}
\end{figure}
\begin{figure}[h!]
    
    \includegraphics[width = 1\linewidth]{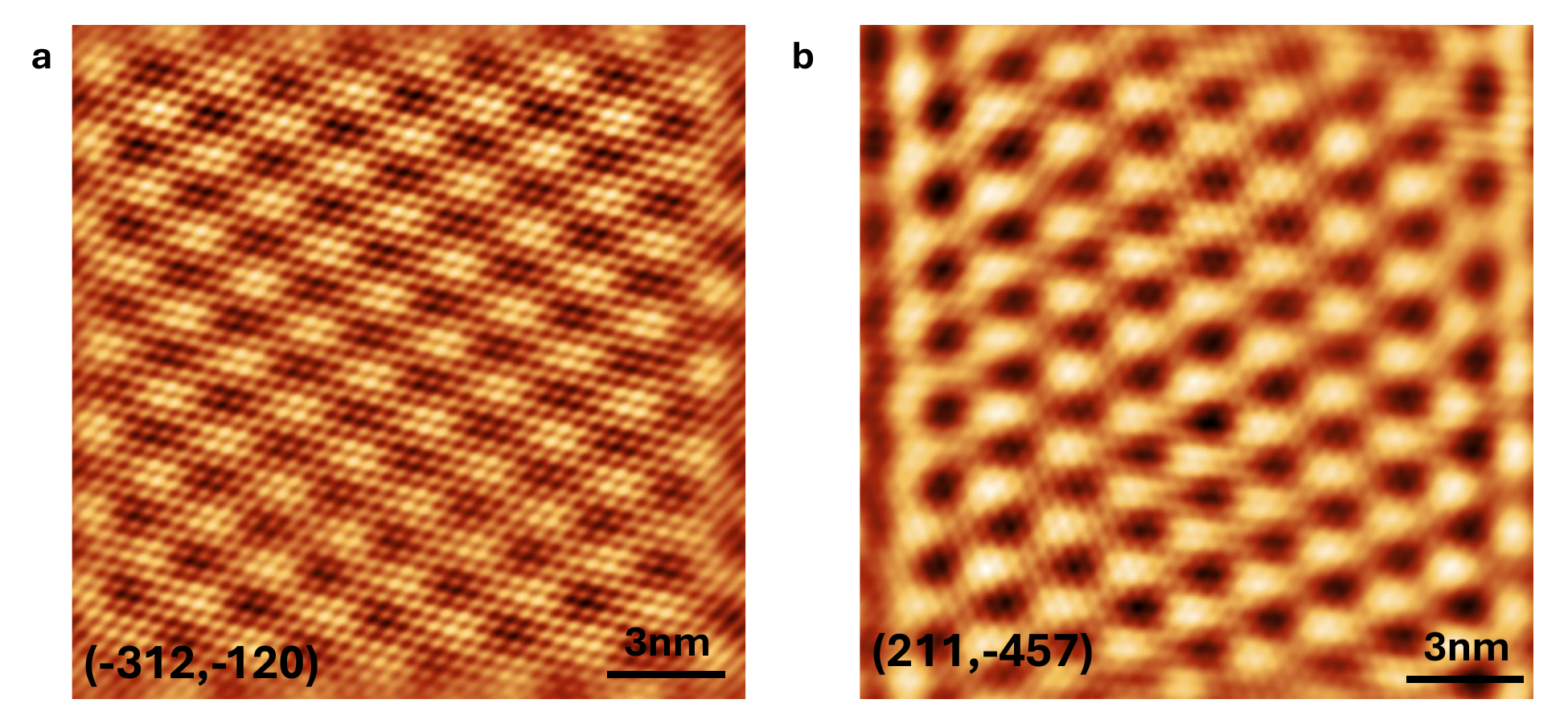}
    
    \caption {(a) An STM topograph of 10.1 $\times$ 10.1 nm$^2$ area taken at specified coordinates showing atomic resolution along with moiré lattice corresponding to twist angle $\sim$ 14.3$^\circ$. (b) Another STM topograph of 11 $\times$ 11 nm$^2$ area at approx. 500 nm distance from (a) also has the same moiré lattice periodicity.}
\end{figure}
\begin{figure}[h!]
    
    \includegraphics[width = 1\linewidth]{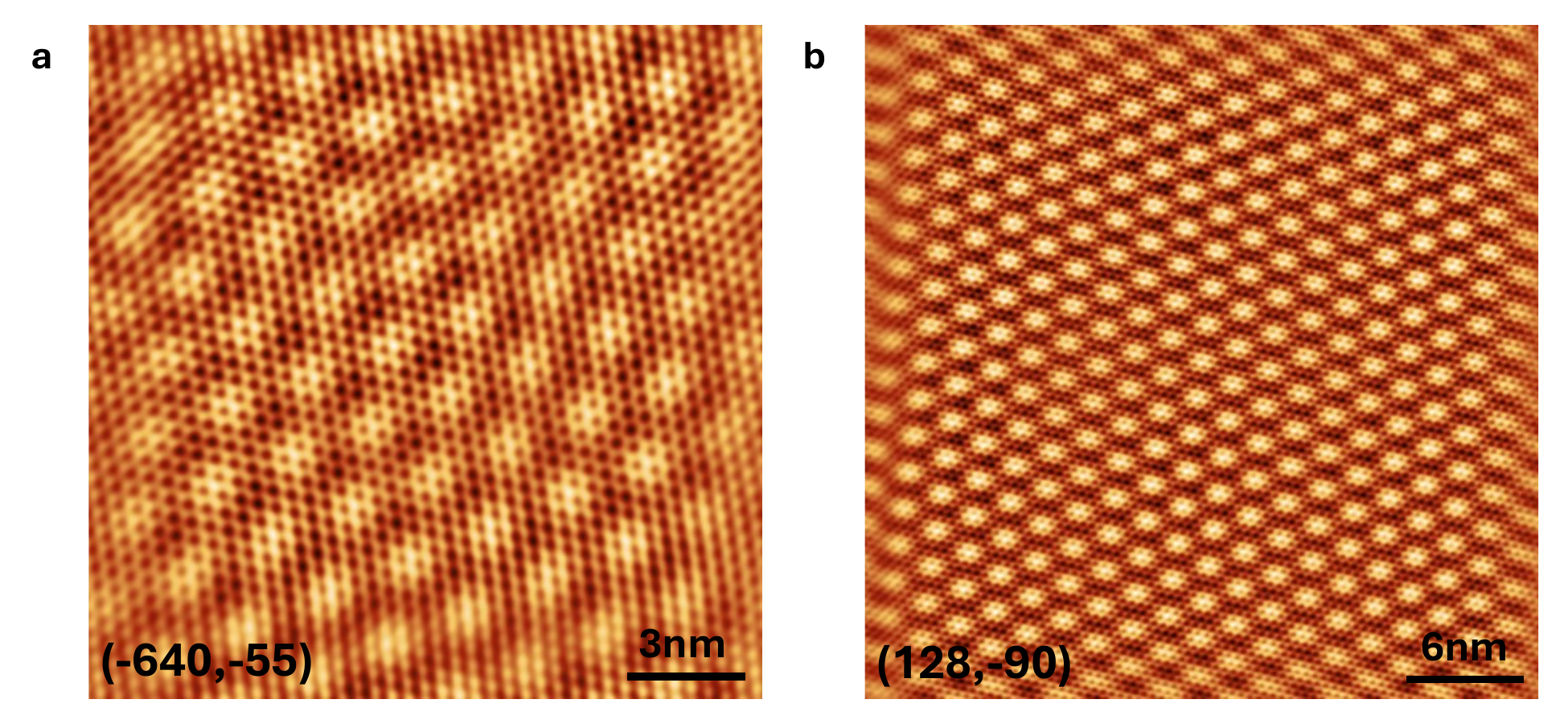}
    
    \caption {(a) An STM topograph of 10.4 $\times$ 10.4 nm$^2$ area taken at specified coordinates showing atomic resolution along with moiré lattice corresponding to twist angle $\sim$ 14$^\circ$. (b) Another STM topograph of 22.4 $\times$ 22.4 nm$^2$ area at approx. 800 nm distance from (a) also has the same moiré lattice periodicity.}
\end{figure}
\clearpage
\newpage
\section*{Raman mapping analysis}
As characterized and described in detail in \cite{gupta_twist-dependent_2020}, the I$_{2D}$/I$_{G}$ peak ratios from Raman spectroscopy are correlated with the twist angle distribution in twisted multilayer graphene films. The uniformity of the intensity ratio between the 2D and G peaks provides qualitative supporting evidence that the twist angle between the layers remains consistent at the micrometer scale.

\begin{figure}[h!]
    
    \includegraphics[width=0.9\linewidth]{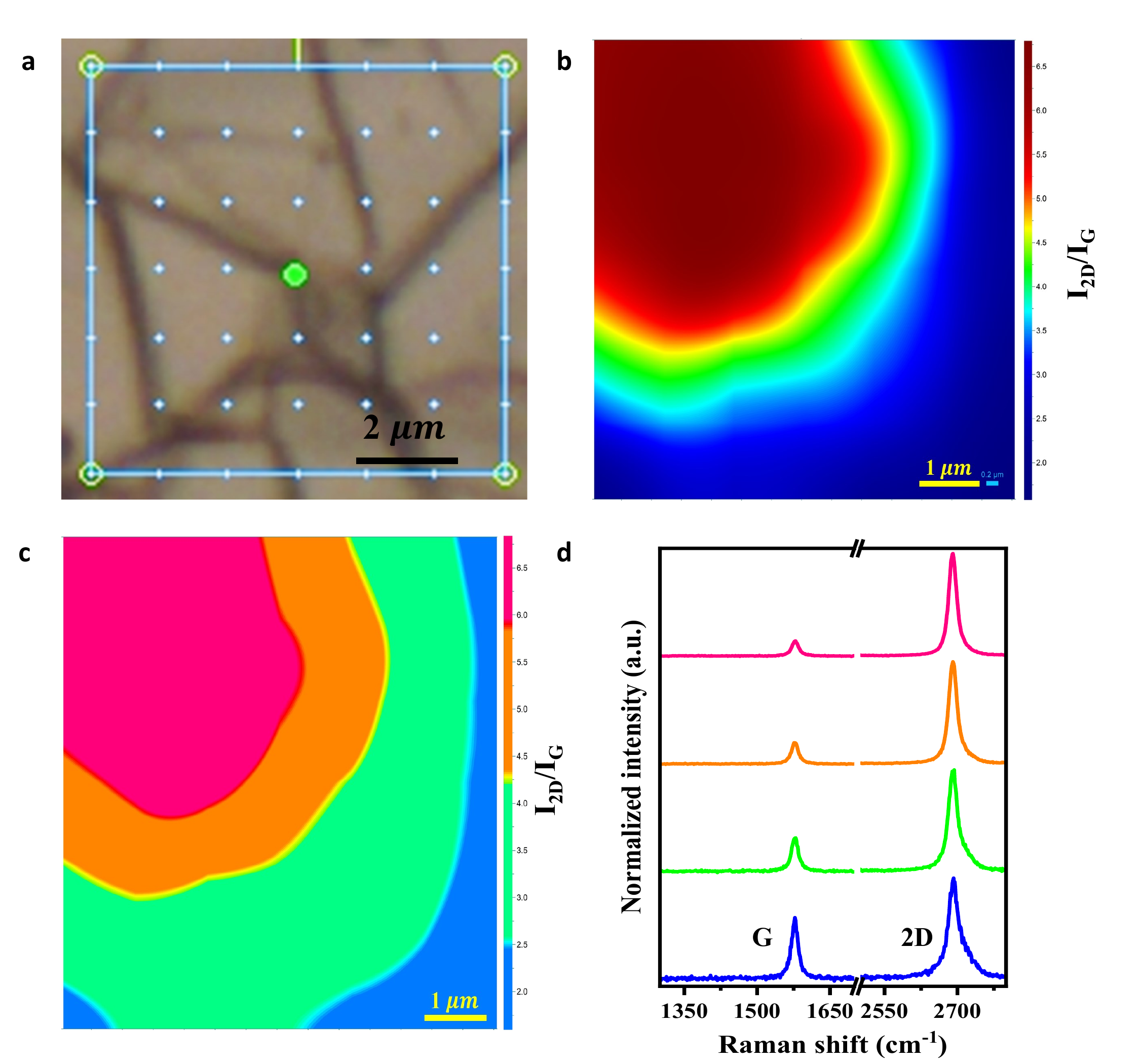}
  
    \caption {(a) Optical microscopy image of one of the regions ($\sim$ 6 $\mu$m $\times$ 6 $\mu$m) of turbostratic graphene. The 7 $\times$ 7 grid (with green spots) indicates regions where Raman spectra were acquired. (b) Corresponding map of I$_{2D}$/I$_{G}$ peak ratios. (c) Processed Raman mapping where a range of I$_{2D}$/I$_{G}$ peak ratios is assigned to a specific color as marked in the color bar. (d) Representative Raman spectra taken from different regions in (c), as indicated by the respective color.}
\end{figure}
\begin{figure}[h!]
   
    \includegraphics[width=\linewidth]{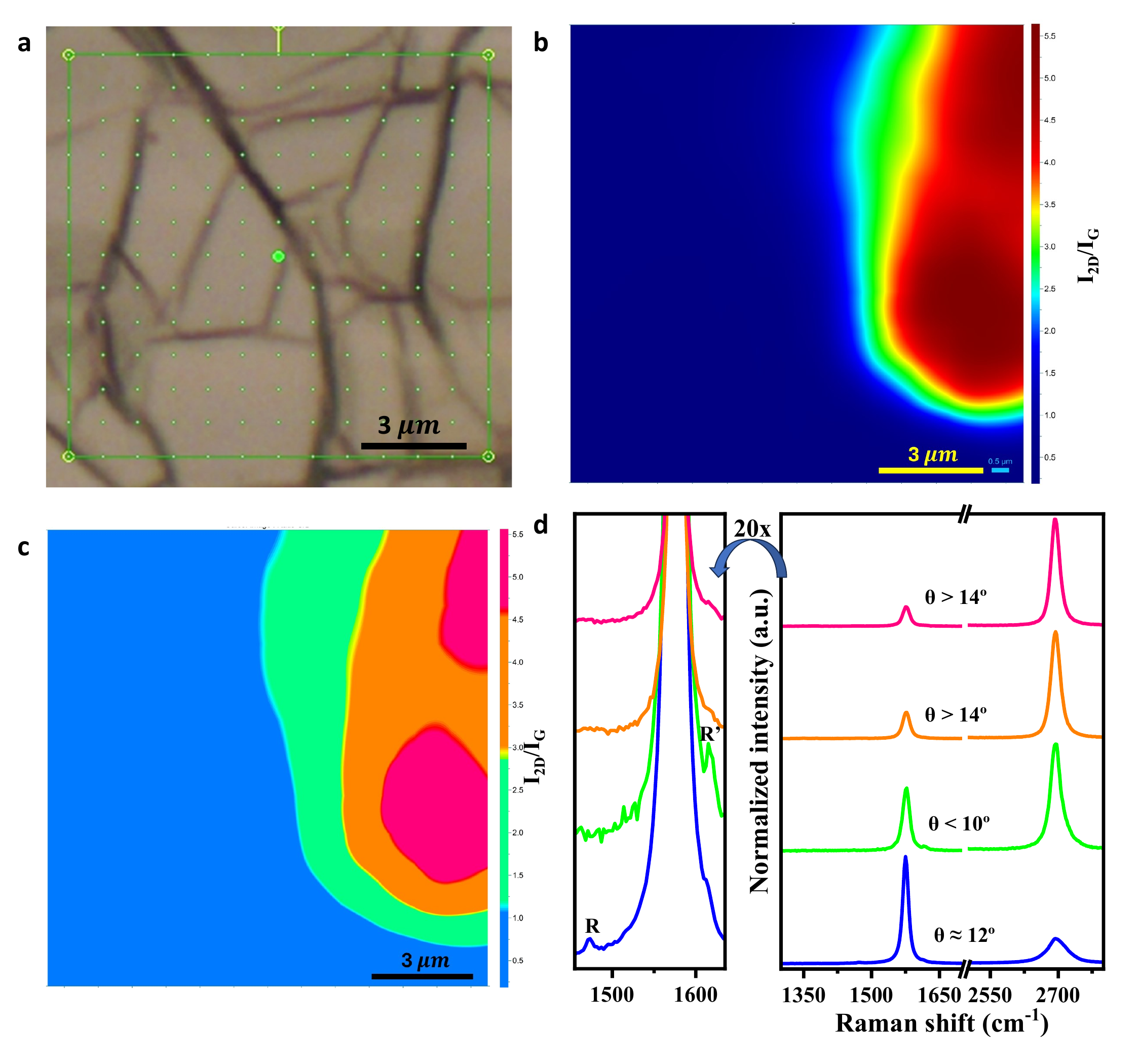}
    
    \caption {(a) Optical microscopy image of one of the regions ($\sim$12 $\mu$m $\times$ 12 $\mu$m) of turbostratic graphene. The 13 $\times$ 13 grid (with green spots) indicates regions where Raman spectra were acquired. (b) Corresponding map of I$_{2D}$/I$_{G}$ peak ratios. (c) Processed Raman mapping where a range of I$_{2D}$/I$_{G}$ peak ratios is assigned to a specific color as marked in the color bar. (d) Representative Raman spectra taken from different regions in (c), as indicated by the respective color.}
\end{figure}
\clearpage
\newpage
\section*{Raman spectroscopy analysis of twist angle distribution}
\begin{figure}[h!]
    \begin{center}
    \includegraphics[width = 1\linewidth]{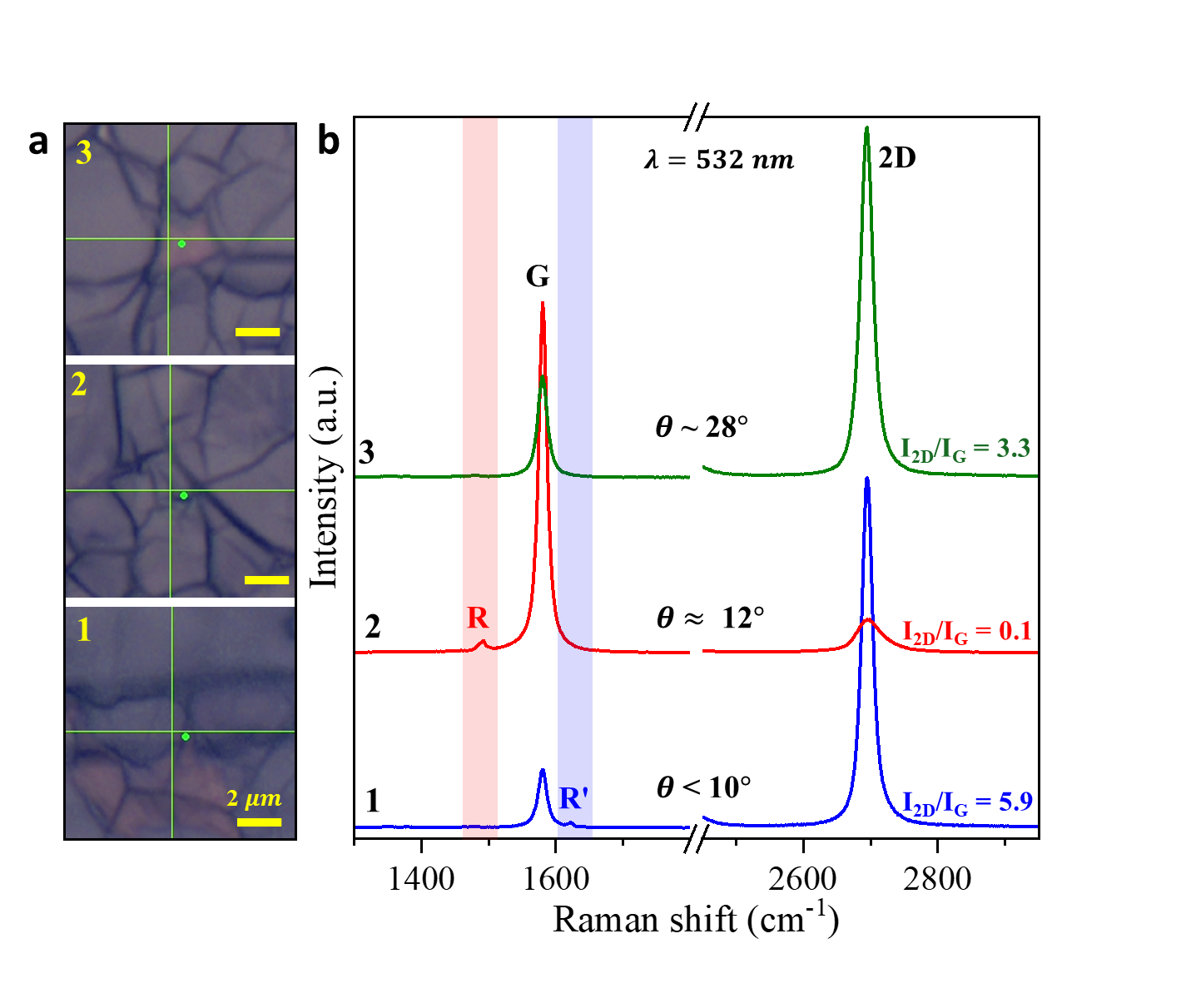}
    \end{center}
    \caption {(a) Optical microscopy images of a few locations of turbostratic graphene with a green dot indicating the location of the laser beam. (b) Raman spectra (532 nm) from these locations, 1 (blue), 2 (red) and 3 (green). The vertical color bands stand for the $R$ (red) and $R^{'}$ (blue) peaks. The twist angle ($\theta$) is a notion based on comparisons made with spectral features in reference \cite{sun_hetero-site_2021}.}
\end{figure}
In Figure S7(a), we show optical microscope images of the regions (marked by the green dots) of a graphene film where the Raman spectroscopy experiments were carried out. The corresponding Raman spectra, acquired using a 532 nm laser, are shown in Figure S7(b). The Raman spectra of all three marked regions exhibit prominent $G$ and $2D$ peaks, located at approximately 1580 cm$^{-1}$ and 2695 cm$^{-1}$, respectively. The spectrum labeled as 1 closely resembles the Raman signature similar to bilayer graphene with an interlayer twist angle of less than 10$^o$\cite{sun_hetero-site_2021}. This spectrum features the rotational peak termed as $R^{'}$ peak around 1622 cm$^{-1}$. However, unlike twisted bilayer graphene with a twist angle below 10$^o$, where the intensity ratio of the 2D to G peaks (I$_{2D}$/I$_G$) is typically less than 1\cite{sun_hetero-site_2021, kim_raman_2012}, this region of turbostratic graphene demonstrates higher I$_{2D}$/I$_G$ ratio due to the presence of more than two layers\cite{mogera_highly_2015}. In the presented spectrum, I$_{2D}$/I$_G$ is measured to be 5.9.

The spectrum labeled as 2 corresponds to the unique stacking arrangement in graphene layers known as the G-enhancement. This phenomenon occurs when the laser excitation energy matches the energy difference between the conduction and valence Van Hove singularities\cite{kim_raman_2012}. For bilayer graphene with a twist angle of approximately 12$^o$, this strong resonance is observed at a laser wavelength of 532 nm (energy = 2.33 eV)\cite{sun_hetero-site_2021}. In the presented spectrum, the G peak is strongly enhanced compared to the 2D peak, with both peaks exhibiting a single Lorentzian profile. The presented spectrum shows I$_{2D}$/I$_G$ = 0.1. This observation is consistent with our STM image analysis which also revealed a dominant existence of the twist angle of 12$^o$. Furthermore, in spectrum 2, an additional rotational peak, termed as the $R$ peak, is observed around 1490 cm$^{-1}$, with an enhanced intensity.

The spectrum labeled as 3 represents the Raman signature similar to bilayer graphene with an interlayer twist angle $\sim$ 28$^o$. In this case, the 2D peak intensity exceeds the G peak. The presented spectrum reveals $I_{2D}/I_G$. Furthermore, the characteristic rotational peaks observed at smaller twist angles, the $R^{'}$ peak (associated with twist angles below 10$^o$) and the R peak (associated with G-enhancement) are not prominently resolved in this spectrum. Therefore, in agreement with the STM results, the Raman measurements confirm the existence of a large range of twist angles between multiple layers of graphene sheets in our turbostratic graphene films.
\clearpage
\newpage
\section*{STM topograph near wrinkled region}

\begin{figure}[h!]
   
    \includegraphics[width = 1\linewidth]{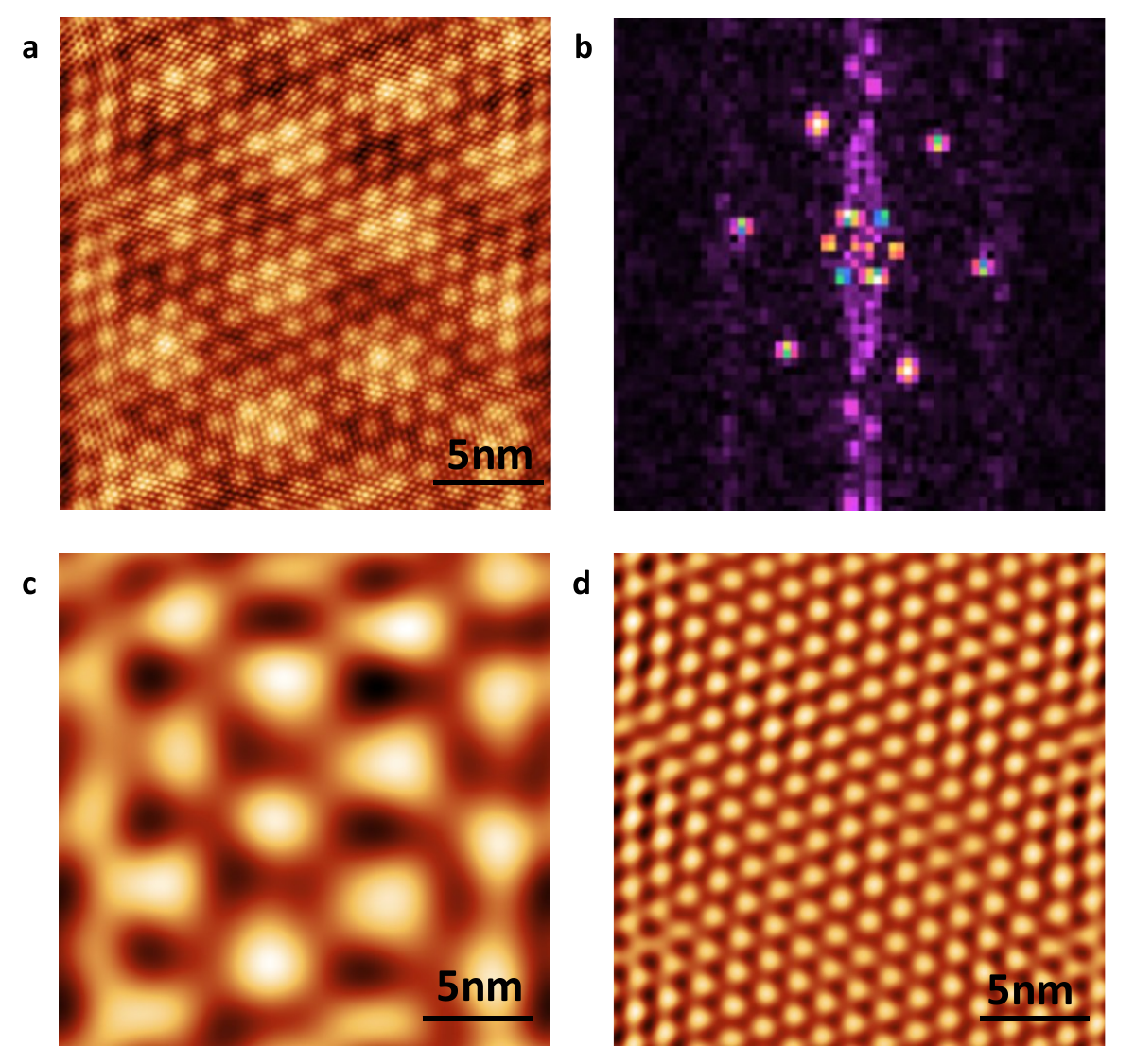}
   
    \caption {(a) An STM topograph of 20.1 $\times$ 20.1 nm$^2$ area taken at a distance $\sim$ 150 nm from wrinkled region as shown in Figure 4(a) in the manuscript showing two different moiré periodicities. (b) Fast Fourier Transform of the topographic image shown in (a). (c, d) Two different moiré superlattices with twist angles $\sim$ 3.1$^\circ$ and 11.2$^\circ$, respectively, extracted by FFT filtering.}
\end{figure}
\clearpage
\newpage

\section*{Temperature Dependence of Conductance Spectra with PLL Features}
\begin{figure}[h!]
    
    \includegraphics[width=1\textwidth]{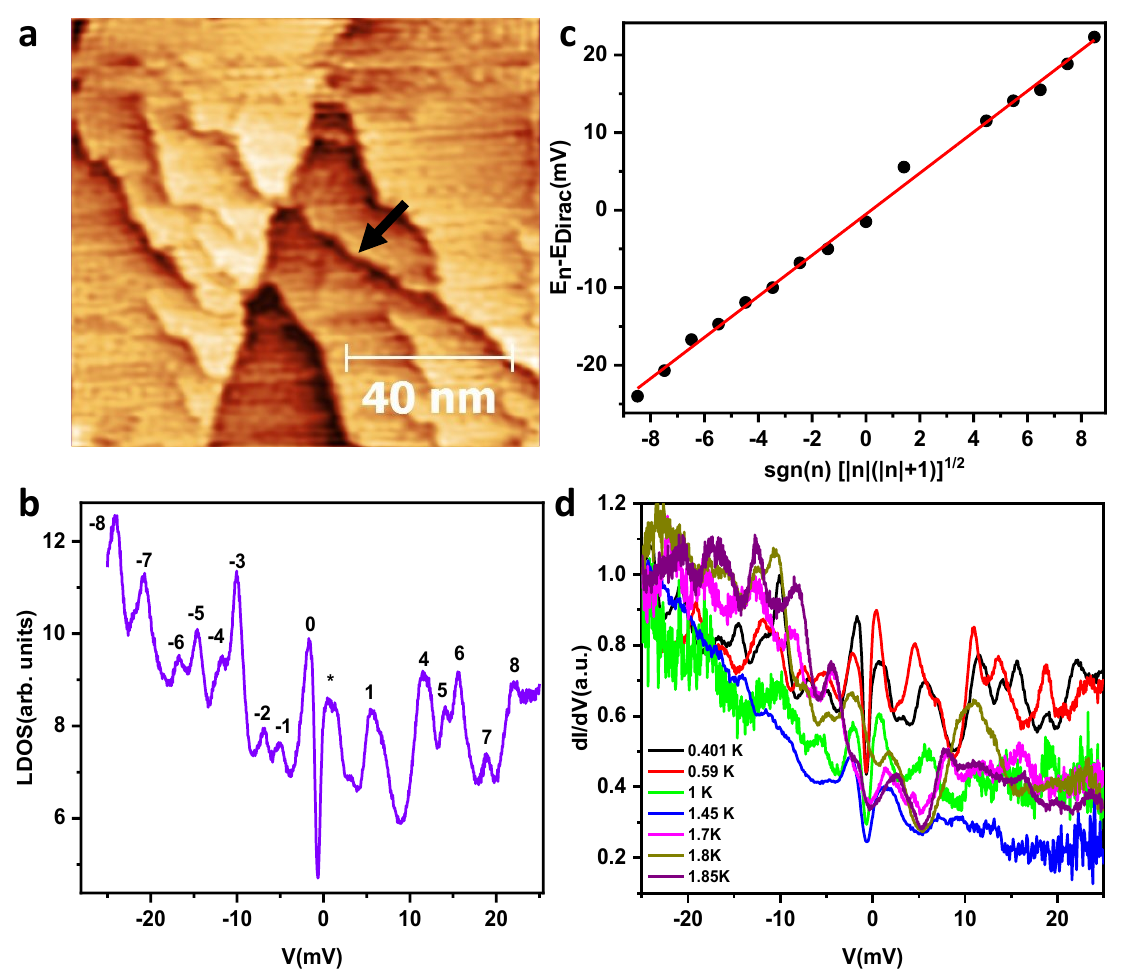}
    
    \caption {(a) An STM Topograph of 111 $\times$ 111 nm$^2$ area when STS was performed. (b) $dI/dV$ characteristics at the marked region at 350 mK. (c) Scaling of E$_n$-E$_{Dirac}$ with $sgn(n)\sqrt{(|n|(|n|+1))}$. (d) Temperature evolution of PLL peaks in $dI/dV$ characteristics.}
\end{figure}

\end{document}